\documentclass[aps,prd,superscriptaddress,showpacs,nofootinbib,floatfix,twocolumn]{revtex4}

\usepackage{graphicx}   

\usepackage{amsmath}

\begin{document}


\title{Measurement of Transverse Single-Spin Asymmetries 
for $J/\psi$ Production \\
in Polarized $p+p$ Collisions at $\sqrt{s} = 200$ GeV}


\newcommand{\abilene}{Abilene Christian University, Abilene, Texas 79699, USA}
\newcommand{\acadsin}{Institute of Physics, Academia Sinica, Taipei 11529, Taiwan}
\newcommand{\banaras}{Department of Physics, Banaras Hindu University, Varanasi 221005, India}
\newcommand{\barc}{Bhabha Atomic Research Centre, Bombay 400 085, India}
\newcommand{\bnlcoll}{Collider-Accelerator Department, Brookhaven National Laboratory, Upton, New York 11973-5000, USA}
\newcommand{\bnlphys}{Physics Department, Brookhaven National Laboratory, Upton, New York 11973-5000, USA}
\newcommand{\caucr}{University of California - Riverside, Riverside, California 92521, USA}
\newcommand{\charlesczech}{Charles University, Ovocn\'{y} trh 5, Praha 1, 116 36, Prague, Czech Republic}
\newcommand{\chonbuk}{Chonbuk National University, Jeonju, 561-756, Korea}
\newcommand{\ciae}{China Institute of Atomic Energy (CIAE), Beijing, People's Republic of China}
\newcommand{\cns}{Center for Nuclear Study, Graduate School of Science, University of Tokyo, 7-3-1 Hongo, Bunkyo, Tokyo 113-0033, Japan}
\newcommand{\colorado}{University of Colorado, Boulder, Colorado 80309, USA}
\newcommand{\columbia}{Columbia University, New York, New York 10027 and Nevis Laboratories, Irvington, New York 10533, USA}
\newcommand{\czechtech}{Czech Technical University, Zikova 4, 166 36 Prague 6, Czech Republic}
\newcommand{\dapnia}{Dapnia, CEA Saclay, F-91191, Gif-sur-Yvette, France}
\newcommand{\debrecen}{Debrecen University, H-4010 Debrecen, Egyetem t{\'e}r 1, Hungary}
\newcommand{\elte}{ELTE, E{\"o}tv{\"o}s Lor{\'a}nd University, H - 1117 Budapest, P{\'a}zm{\'a}ny P. s. 1/A, Hungary}
\newcommand{\ewha}{Ewha Womans University, Seoul 120-750, Korea}
\newcommand{\fit}{Florida Institute of Technology, Melbourne, Florida 32901, USA}
\newcommand{\fsu}{Florida State University, Tallahassee, Florida 32306, USA}
\newcommand{\gsu}{Georgia State University, Atlanta, Georgia 30303, USA}
\newcommand{\hiroshima}{Hiroshima University, Kagamiyama, Higashi-Hiroshima 739-8526, Japan}
\newcommand{\ihepprot}{IHEP Protvino, State Research Center of Russian Federation, Institute for High Energy Physics, Protvino, 142281, Russia}
\newcommand{\illuiuc}{University of Illinois at Urbana-Champaign, Urbana, Illinois 61801, USA}
\newcommand{\instpasczech}{Institute of Physics, Academy of Sciences of the Czech Republic, Na Slovance 2, 182 21 Prague 8, Czech Republic}
\newcommand{\isu}{Iowa State University, Ames, Iowa 50011, USA}
\newcommand{\jinrdubna}{Joint Institute for Nuclear Research, 141980 Dubna, Moscow Region, Russia}
\newcommand{\jyvaskyla}{Helsinki Institute of Physics and University of Jyv{\"a}skyl{\"a}, P.O.Box 35, FI-40014 Jyv{\"a}skyl{\"a}, Finland}
\newcommand{\kek}{KEK, High Energy Accelerator Research Organization, Tsukuba, Ibaraki 305-0801, Japan}
\newcommand{\kfki}{KFKI Research Institute for Particle and Nuclear Physics of the Hungarian Academy of Sciences (MTA KFKI RMKI), H-1525 Budapest 114, POBox 49, Budapest, Hungary}
\newcommand{\korea}{Korea University, Seoul, 136-701, Korea}
\newcommand{\kurchatov}{Russian Research Center ``Kurchatov Institute", Moscow, Russia}
\newcommand{\kyoto}{Kyoto University, Kyoto 606-8502, Japan}
\newcommand{\labllr}{Laboratoire Leprince-Ringuet, Ecole Polytechnique, CNRS-IN2P3, Route de Saclay, F-91128, Palaiseau, France}
\newcommand{\lawllnl}{Lawrence Livermore National Laboratory, Livermore, California 94550, USA}
\newcommand{\losalamos}{Los Alamos National Laboratory, Los Alamos, New Mexico 87545, USA}
\newcommand{\lpc}{LPC, Universit{\'e} Blaise Pascal, CNRS-IN2P3, Clermont-Fd, 63177 Aubiere Cedex, France}
\newcommand{\lund}{Department of Physics, Lund University, Box 118, SE-221 00 Lund, Sweden}
\newcommand{\maryland}{University of Maryland, College Park, Maryland 20742, USA}
\newcommand{\mass}{Department of Physics, University of Massachusetts, Amherst, Massachusetts 01003-9337, USA }
\newcommand{\muenster}{Institut fur Kernphysik, University of Muenster, D-48149 Muenster, Germany}
\newcommand{\muhlenberg}{Muhlenberg College, Allentown, Pennsylvania 18104-5586, USA}
\newcommand{\myongji}{Myongji University, Yongin, Kyonggido 449-728, Korea}
\newcommand{\nagasaki}{Nagasaki Institute of Applied Science, Nagasaki-shi, Nagasaki 851-0193, Japan}
\newcommand{\newmex}{University of New Mexico, Albuquerque, New Mexico 87131, USA }
\newcommand{\nmsu}{New Mexico State University, Las Cruces, New Mexico 88003, USA}
\newcommand{\ornl}{Oak Ridge National Laboratory, Oak Ridge, Tennessee 37831, USA}
\newcommand{\orsay}{IPN-Orsay, Universite Paris Sud, CNRS-IN2P3, BP1, F-91406, Orsay, France}
\newcommand{\peking}{Peking University, Beijing, People's Republic of China}
\newcommand{\pnpi}{PNPI, Petersburg Nuclear Physics Institute, Gatchina, Leningrad region, 188300, Russia}
\newcommand{\riken}{RIKEN Nishina Center for Accelerator-Based Science, Wako, Saitama 351-0198, Japan}
\newcommand{\rikjrbrc}{RIKEN BNL Research Center, Brookhaven National Laboratory, Upton, New York 11973-5000, USA}
\newcommand{\rikkyo}{Physics Department, Rikkyo University, 3-34-1 Nishi-Ikebukuro, Toshima, Tokyo 171-8501, Japan}
\newcommand{\saispbstu}{Saint Petersburg State Polytechnic University, St. Petersburg, Russia}
\newcommand{\saopaulo}{Universidade de S{\~a}o Paulo, Instituto de F\'{\i}sica, Caixa Postal 66318, S{\~a}o Paulo CEP05315-970, Brazil}
\newcommand{\seoulnat}{Seoul National University, Seoul, Korea}
\newcommand{\stonybrkc}{Chemistry Department, Stony Brook University, SUNY, Stony Brook, New York 11794-3400, USA}
\newcommand{\stonycrkp}{Department of Physics and Astronomy, Stony Brook University, SUNY, Stony Brook, New York 11794-3400, USA}
\newcommand{\subatech}{SUBATECH (Ecole des Mines de Nantes, CNRS-IN2P3, Universit{\'e} de Nantes) BP 20722 - 44307, Nantes, France}
\newcommand{\tenn}{University of Tennessee, Knoxville, Tennessee 37996, USA}
\newcommand{\titech}{Department of Physics, Tokyo Institute of Technology, Oh-okayama, Meguro, Tokyo 152-8551, Japan}
\newcommand{\tsukuba}{Institute of Physics, University of Tsukuba, Tsukuba, Ibaraki 305, Japan}
\newcommand{\vandy}{Vanderbilt University, Nashville, Tennessee 37235, USA}
\newcommand{\waseda}{Waseda University, Advanced Research Institute for Science and Engineering, 17 Kikui-cho, Shinjuku-ku, Tokyo 162-0044, Japan}
\newcommand{\weizmann}{Weizmann Institute, Rehovot 76100, Israel}
\newcommand{\yonsei}{Yonsei University, IPAP, Seoul 120-749, Korea}
\affiliation{\abilene}
\affiliation{\acadsin}
\affiliation{\banaras}
\affiliation{\barc}
\affiliation{\bnlcoll}
\affiliation{\bnlphys}
\affiliation{\caucr}
\affiliation{\charlesczech}
\affiliation{\chonbuk}
\affiliation{\ciae}
\affiliation{\cns}
\affiliation{\colorado}
\affiliation{\columbia}
\affiliation{\czechtech}
\affiliation{\dapnia}
\affiliation{\debrecen}
\affiliation{\elte}
\affiliation{\ewha}
\affiliation{\fit}
\affiliation{\fsu}
\affiliation{\gsu}
\affiliation{\hiroshima}
\affiliation{\ihepprot}
\affiliation{\illuiuc}
\affiliation{\instpasczech}
\affiliation{\isu}
\affiliation{\jinrdubna}
\affiliation{\jyvaskyla}
\affiliation{\kek}
\affiliation{\kfki}
\affiliation{\korea}
\affiliation{\kurchatov}
\affiliation{\kyoto}
\affiliation{\labllr}
\affiliation{\lawllnl}
\affiliation{\losalamos}
\affiliation{\lpc}
\affiliation{\lund}
\affiliation{\maryland}
\affiliation{\mass}
\affiliation{\muenster}
\affiliation{\muhlenberg}
\affiliation{\myongji}
\affiliation{\nagasaki}
\affiliation{\newmex}
\affiliation{\nmsu}
\affiliation{\ornl}
\affiliation{\orsay}
\affiliation{\peking}
\affiliation{\pnpi}
\affiliation{\riken}
\affiliation{\rikjrbrc}
\affiliation{\rikkyo}
\affiliation{\saispbstu}
\affiliation{\saopaulo}
\affiliation{\seoulnat}
\affiliation{\stonybrkc}
\affiliation{\stonycrkp}
\affiliation{\subatech}
\affiliation{\tenn}
\affiliation{\titech}
\affiliation{\tsukuba}
\affiliation{\vandy}
\affiliation{\waseda}
\affiliation{\weizmann}
\affiliation{\yonsei}
\author{A.~Adare} \affiliation{\colorado}
\author{S.~Afanasiev} \affiliation{\jinrdubna}
\author{C.~Aidala} \affiliation{\mass}
\author{N.N.~Ajitanand} \affiliation{\stonybrkc}
\author{Y.~Akiba} \affiliation{\riken} \affiliation{\rikjrbrc}
\author{H.~Al-Bataineh} \affiliation{\nmsu}
\author{J.~Alexander} \affiliation{\stonybrkc}
\author{H.~Al-Ta'ani} \affiliation{\nmsu}
\author{A.~Angerami} \affiliation{\columbia}
\author{K.~Aoki} \affiliation{\kyoto} \affiliation{\riken}
% ...
\collaboration{PHENIX Collaboration} \noaffiliation

\begin{abstract}



We report the first measurement of transverse single-spin 
asymmetries in $J/\psi$ production from transversely polarized $p+p$ 
collisions at $\sqrt{s} = 200$~GeV with data taken by the PHENIX 
experiment in 2006 and 2008.  The measurement was performed over the 
rapidity ranges $1.2 < |y| < 2.2$ and $ |y| < 0.35$ for transverse 
momenta up to 6~GeV/$c$.  $J/\psi$ production at RHIC is dominated by 
processes involving initial-state gluons, and transverse single-spin 
asymmetries of the $J/\psi$ can provide access to gluon dynamics 
within the nucleon.  Such asymmetries may also shed light on the 
long-standing question in QCD of the $J/\psi$ production mechanism.  
Asymmetries were obtained as a function of $J/\psi$ transverse 
momentum and Feynman-$x$, with a value of 
$-0.086~\pm~0.026^{\rm stat}~\pm~0.003^{\rm syst}$ in the forward 
region.  This result suggests possible nonzero trigluon correlation 
functions in transversely polarized protons and, if well defined 
in this reaction, a nonzero gluon Sivers distribution function.

\end{abstract}

\pacs{14.20.Dh, 13.88.+e, 13.85.Ni, 14.40.Pq}

\maketitle




\section{Introduction\label{sec:Introduction}}

The transverse single-spin asymmetry (SSA) quantifies the asymmetry of 
particle production relative to the plane defined by the transverse 
spin axis and the momentum direction of a polarized hadron.  SSAs 
have come to be recognized as a means of accessing QCD dynamics, 
both within initial-state hadrons and in the process of 
hadronization from partons.  Large azimuthal transverse single-spin 
asymmetries of up to $\sim 40$\% were first observed at large 
Feynman-$x$ ($x_F = 2p_L/\sqrt{s}$, where $p_L$ is the momentum 
along the beam direction) in pion production from transversely 
polarized $p+p$ collisions at $\sqrt{s}=4.9$~GeV in 1976 
\cite{Klem:1976ui}, contrary to theoretical expectations at the 
time~\cite{Kane:1978nd}, and subsequently observed in hadronic 
collisions over a range of energies extending up to 
$\sqrt{s}=200$~GeV \cite{Allgower:2002qi, Antille:1980th, 
Adams:1991rw, Adams:1991cs, Arsene:2008mi, Adams:2003fx, 
Abelev:2008qb}.  In recent years numerous measurements of transverse 
SSAs have been performed in semi-inclusive deep-inelastic scattering 
(SIDIS) off a transversely polarized proton or deuteron target as 
well~\cite{Airapetian:2004tw, Alexakhin:2005iw, Ageev:2006da, 
Airapetian:2008sk, Alekseev:2008dn, Airapetian:2009ti}.  In order to 
describe the large transverse SSAs observed, two approaches have 
been developed since the 1990s, after early pioneering work by 
Efremov and Teryaev \cite{Efremov:1981sh, Efremov:1984ip}.  One 
approach requires higher-twist contributions in the collinear 
factorization scheme, i.e.~the exchange of a gluon between one of 
the partons taking part in the hard scattering and the color field 
of either an initial- or final-state hadron.  This was first 
proposed by Qiu and Sterman for gluon exchange in the initial state 
\cite{Qiu:1998ia} and by Kanazawa and Koike for exchange in the 
final state \cite{Kanazawa:2000hz}.  Gluon exchange in either the 
initial or final state leads to terms including multiparton 
correlation functions, which can generate a nonzero SSA.  The other 
approach utilizes parton distribution functions and/or fragmentation 
functions that are unintegrated in the partonic transverse momentum, 
$k_T$; these functions are generally known as 
transverse-momentum-dependent distributions (TMDs).  These two 
approaches have different but overlapping kinematic regimes of 
applicability and have been shown to correspond exactly in their 
region of overlap \cite{Ji:2006ub}.

While higher-twist parton-nucleon spin-momentum correlation 
functions and TMDs were born within the nucleon structure community, 
as understanding of them matures, their implications for other areas 
of QCD are starting to be realized.  $J/\psi$ production has been 
extensively studied over the last decades, but the details of the 
production mechanism remain an open question (see 
ref.~\cite{Lansberg:2006dh} for a comprehensive review), and an 
additional complication is that approximately 30\%--40\% of the 
measured $J/\psi$ mesons in hadronic collisions are produced 
indirectly from feed-down from $\psi'$ and 
$\chi_c$~\cite{Abt:2002vq}.  It was proposed in 2008 by Yuan 
\cite{Yuan:2008vn} that within the framework of nonrelativistic QCD 
(NRQCD) \cite{Bodwin:1994jh}, the transverse SSA of $J/\psi$ 
production can be sensitive to the $J/\psi$ production mechanism, 
assuming a nonzero gluon Sivers function \cite{Sivers:1989cc}, 
which is a TMD that describes the correlation between the transverse 
spin of the proton and the $k_T$ of the partons within it.  
Specifically, Yuan predicts that a nonzero gluon Sivers function 
will produce a finite transverse SSA for color-singlet $J/\psi$ 
production~\cite{Baier:1981uk} in $p+p$ collisions, but the 
asymmetry should vanish for color-octet 
production~\cite{Bodwin:1994jh} in $p+p$ due to cancelation between 
initial- and final-state effects, while a nonzero asymmetry for 
$J/\psi$ production in SIDIS is only expected within the color-octet 
model.  It should be noted that the relationship between the 
transverse SSA and the production mechanism is not quite as simple 
in the collinear higher-twist approach, with partial but not full 
cancelation of terms~\cite{Kang:2010pr} in the cases where the 
asymmetry uniformly vanishes in the TMD approach presented by Yuan.  
Another important point to note regarding the TMD as compared to the 
collinear, higher-twist approach is that very recent theoretical 
work~\cite{Rogers:2010dm} suggests that factorization of hard 
processes in perturbative QCD (pQCD) into 
transverse-momentum-dependent distribution and fragmentation 
functions convoluted with partonic hard-scattering cross sections is 
not valid for processes involving more than two hadrons.  Thus, in 
the process $p+p \rightarrow J/\psi + X$ a gluon Sivers function may 
not be well defined; however, the definition within a factorized 
pQCD framework of the corresponding trigluon correlation functions 
in the collinear, higher-twist approach is believed to be valid.

Measurements of heavy flavor transverse SSAs in $p+p$ 
collisions are of interest because they serve to isolate gluon 
dynamics within the nucleon, with heavy quarks being dominantly 
produced via gluon-gluon interactions.  Very little is presently 
known about trigluon correlation functions or gluon TMDs.  
Measurements of the transverse momentum ($p_T$) spectrum of 
bottomonium production have been used to extract the 
$k_T$-unintegrated distribution of gluons in an unpolarized proton 
\cite{Kulesza:2003wi}.  Similarly, measurements of open heavy flavor 
transverse SSAs have been proposed as a means to isolate gluon TMDs 
and/or their corresponding twist-three gluon correlation functions 
in polarized protons~\cite{Anselmino:2004nk, Kang:2008ih, 
Yuan:2008it}.  A previous PHENIX measurement of the transverse SSA in 
neutral pion production at midrapidity \cite{Adler:2005in} as well 
as measurements by the COMPASS collaboration of the SSA in 
semi-inclusive charged hadron production~\cite{Alexakhin:2005iw}, 
both consistent with zero, have been used to provide initial 
constraints on the gluon Sivers function and suggest that it is 
small~\cite{Anselmino:2006yq, Brodsky:2006ha}, but further data 
relevant to gluon TMDs are greatly needed.

In this paper the first measurement of transverse SSAs in $J/\psi$ 
production is presented.  The data were taken by the PHENIX 
experiment at the Relativistic Heavy Ion Collider (RHIC) during the 
2006 and 2008 polarized proton runs at $\sqrt{s}=200$~GeV.  The 
$p_T$ and $x_F$ dependencies are studied, for rapidity regions of 
$-2.2 < y < -1.2$, $|y|<0.35$, and $1.2 < y < 2.2$, and $p_T$ up to 
6 GeV/$c$.

\section{Analysis\label{sec:Method-main}}

\subsection{Measuring Transverse Single-Spin Asymmetries
\label{subsec:SSA}}

Transverse single-spin asymmetries lead to modulations of the cross 
section in the azimuthal angle due to the projection of the 
polarization vector into the direction of the produced particle.  
Our measurement is performed in two separate hemispheres referred to 
as `left' and `right' where left is defined as the axis which forms 
a right-handed coordinate system with the beam momentum vector and 
one of the spin orientations, denoted as $\uparrow$.  For a vector 
$\vec{S}$ in the direction of the $\uparrow$ spin and beam momentum 
$\vec{P}$, left is defined as $\vec{p} \cdot 
\left(\vec{S}\times\vec{P}\right)>0$ with $\vec{p}$ being the 
momentum vector of the outgoing particle.

The left-right transverse SSA can be extracted using 
Eq.~(\ref{eq:AN-sigma}).  This equation applies to particle yields 
observed to the left side of the polarized beam.

\begin{equation}
\label{eq:AN-sigma} A_{N} = \frac{f}{\mathcal{P}}
\frac{{(\sigma^\uparrow-\sigma^\downarrow)}}
{{(\sigma^\uparrow+\sigma^\downarrow)}},
\end{equation}
where $\sigma^\uparrow (\sigma^\downarrow)$ represents the 
production cross section with beam polarized in the $\uparrow$ 
($\downarrow$) direction, integrated over the left hemisphere, and 
$\mathcal{P}$ is the beam polarization.  An overall minus sign is 
required for $A_{N}$ on the right side of the polarized beam.

The geometric scale factor $f$ corrects for the convolution of an 
azimuthal asymmetry with detector acceptance.  For a sinusoidal 
asymmetry, as generated by the Sivers function, the factor becomes
\begin{equation}
f = \left(\frac{\int^{\pi}_{0}  \varepsilon(\phi) \sin\phi d\phi}{\int^{\pi}_{0} \varepsilon(\phi) d\phi}\right)^{-1},
\end{equation}
where $\phi$ is the azimuthal angle between the outgoing particle 
and the proton spin, and $\varepsilon(\phi)$ is the efficiency for 
detecting a $J/\psi$ at a given $\phi$.  The limits of integration 
correspond to the hemisphere in which the measurement is being made.  
It should be noted that even for a detector with full azimuthal 
coverage the factor $f$ is not unity because we are not determining 
the amplitude of the modulation directly but measuring an asymmetry 
which is integrated over entire hemispheres in azimuth.

Both proton beams in RHIC were polarized.  In order to derive 
single-spin asymmetries, we sum over polarization direction states 
in one of the two beams: 
\begin{eqnarray}
\sigma^{\uparrow}   &\equiv& \sigma^{\uparrow\uparrow}+\sigma^{\uparrow\downarrow},    \\
\sigma^{\downarrow} &\equiv& \sigma^{\downarrow\uparrow}+\sigma^{\downarrow\downarrow},
\end{eqnarray}
where $\sigma$ again indicates the $\phi$-dependent cross 
section integrated over one hemisphere, and $\uparrow$ and 
$\downarrow$ represent the spin orientations of the two beams.  
Differences in luminosities can lead to false asymmetries; see 
Section~\ref{subsec:Method-central}.  We perform the analysis 
separately for each beam.  Up to 111 out of a possible 120 
bunches were filled in each RHIC ring, with 106~ns between 
bunches.  Four preset spin patterns with approximately equal 
numbers of bunches polarized in opposite directions were 
alternated during the transverse running periods to minimize 
systematic effects from the injection process.  Stores were 
nominally held for eight hours.  The spin orientation at the 
PHENIX interaction point was maintained at the default 
vertical orientation in the RHIC ring for 2008 data taking and 
rotated to radial orientation for 2006.

Explicitly including the spin orientations of both beams, 
Eq.~(\ref{eq:AN-sigma}) can be rewritten as:
\begin{eqnarray}
\label{eq:AN-modified} A_{N} &=& \frac{f}{\mathcal{P}}
\frac{ (\sigma^{\uparrow\uparrow}+\sigma^{\uparrow\downarrow})-
(\sigma^{\downarrow\uparrow}+\sigma^{\downarrow\downarrow}) }
{ (\sigma^{\uparrow\uparrow}+\sigma^{\uparrow\downarrow})+
(\sigma^{\downarrow\uparrow}+\sigma^{\downarrow\downarrow}) } \nonumber \\
&=& \frac{f}{\mathcal{P}}
\frac{ (N^{\uparrow\uparrow}+\mathcal{R}_1N^{\uparrow\downarrow})
-(\mathcal{R}_2N^{\downarrow\uparrow}+\mathcal{R}_3N^{\downarrow\downarrow}) }
{ (N^{\uparrow\uparrow}+\mathcal{R}_1N^{\uparrow\downarrow})
+ (\mathcal{R}_2N^{\downarrow\uparrow}+\mathcal{R}_3N^{\downarrow\downarrow})}, 
\end{eqnarray}
where $N^{\uparrow\uparrow}, N^{\uparrow\downarrow}, 
N^{\downarrow\uparrow}$ and $N^{\downarrow\downarrow}$ are the 
experimental yields in each spin configuration, and 
$\mathcal{R}_1=\mathcal{L}^{\uparrow\uparrow}~/~\mathcal{L}^{\uparrow\downarrow}$, 
$\mathcal{R}_2=\mathcal{L}^{\uparrow\uparrow}~/~\mathcal{L}^{\downarrow\uparrow}$ 
and 
$\mathcal{R}_3=\mathcal{L}^{\uparrow\uparrow}~/~\mathcal{L}^{\downarrow\downarrow}$ 
are ratios of the provided 
luminosities $\mathcal{L}$ in each spin orientation.  The use of a 
single polarization factor assumes that bunch polarizations 
$\mathcal{P}$ are equal regardless of spin orientation in each beam.  
We have also assumed that detector efficiencies are equal for each 
spin orientation, given that the orientations change every 106 or 212~ns.

Beam polarizations at RHIC are measured by two different 
polarimeters, a fast carbon target 
polarimeter~\cite{Nakagawa:2008zzb} for relative polarization 
measurements and a hydrogen jet polarimeter~\cite{Okada:2005gu} for 
an absolute measurement.  The carbon target polarimeter is used to 
measure relative beam polarization store-by-store, and the hydrogen 
jet polarimeter is used for absolute calibration of the beam 
polarization.  During the 2006 run the average transverse beam 
polarizations were
\begin{eqnarray*}
& 0.53\pm 0.02^{\rm syst} & {\rm (clockwise)} \\
& 0.52\pm 0.02^{\rm syst} & {\rm (counterclockwise).}
\end{eqnarray*}

Fill-to-fill variation of beam polarization was $\pm$0.03 (1 
$\sigma$) for the clockwise beam and $\pm$0.04 (1 $\sigma$) for 
the counterclockwise beam.  The labels represent the direction in 
which the beam is circulating when looking from above, and the 
systematic uncertainties are uncorrelated between beams.  There is 
an additional systematic uncertainty of $3.4\%$ correlated between 
the two beams.  The average beam polarizations during the 2008 
data taking were
\begin{eqnarray*}
& 0.48\pm 0.02^{\rm syst} & {\rm (clockwise)} \\
& 0.41\pm 0.02^{\rm syst} & {\rm (counterclockwise)}
\end{eqnarray*}
each with a fill-to-fill variation of $\pm$0.04 (1 $\sigma$), and 
with an additional systematic uncertainty of $3.0\%$ correlated 
between the beams.

For statistically limited measurements it may be impossible to 
measure asymmetries using separate yields based on the spin 
orientation of both beams.  In this case a good approximation can 
be made using
\begin{equation}
\label{eq:AN-lumi} A_{N} = \frac{f}{\mathcal{P}}
\frac{N^{\uparrow}-\mathcal{R}N^{\downarrow}}
{N^{\uparrow} + \mathcal{R}N^{\downarrow}},
\end{equation}
where the spin orientation of the polarized beam is denoted by the 
arrow, and we have a single relative luminosity 
$\mathcal{R}=\frac{\mathcal{L^{\uparrow}}}{\mathcal{L^{\downarrow}}}$.  
Systematic effects which will be discussed in 
Section~\ref{subsec:Method-central} can be reduced by bringing 
$\mathcal{R}$ to a constant value.

To further simplify Eq.~(\ref{eq:AN-lumi}) we can eliminate all 
explicit uses of the relative luminosity and calculate a single 
asymmetry for both the left and right hemispheres using
\begin{equation}
\label{eq:AN-sqrt} A_N = \frac{f^{\prime}}{\mathcal{P}} \frac{{\sqrt{N_L^\uparrow
N_R^\downarrow}-\sqrt{N_L^\downarrow N_R^\uparrow}}}
{{\sqrt{N_L^\uparrow N_R^\downarrow}+\sqrt{N_L^\downarrow
N_R^\uparrow}}}.
\end{equation}
The geometric scale factor
\small
\begin{equation}
f^{\prime} = 2\left(\frac{\int^{\pi}_{0}  \varepsilon(\phi) \sin\phi d\phi}{\int^{\pi}_{0} \varepsilon(\phi) d\phi} - \frac{\int^{2\pi}_{\pi}  \varepsilon(\phi) \sin\phi d\phi}{\int^{2\pi}_{\pi} \varepsilon(\phi) d\phi} \right)^{-1}
\end{equation}
\normalsize
is different from the scale factor in Eq.~(\ref{eq:AN-modified}) and 
Eq.~(\ref{eq:AN-lumi}) because left and right are treated 
simultaneously, leading to differences between the two expressions 
of order $(A_{N})^{3}$.

\subsection{$J/\psi$ Measurements\label{sec:Exp-setup}}

Measurements were carried out by the PHENIX experiment at RHIC, 
where the cross section and polarization of $J/\psi$ mesons in 
$\sqrt{s}=200$~GeV $p+p$ collisions have also been 
measured~\cite{Adare:2006kf,Adare:2009js}.  An overview of the 
PHENIX apparatus can be found in~\cite{Adcox:2003zm}.  At forward 
and backward rapidities, $J/\psi \rightarrow \mu^+\mu^-$ were 
measured with two muon spectrometers~\cite{Akikawa:2003zs}, for 
$1.2 < |y| <2.2$ and $\Delta\phi = 2\pi$, using data recorded in 2006 
and 2008.  At midrapidity, asymmetries were studied via $J/\psi 
\rightarrow e^+e^-$ with the central arm 
spectrometers~\cite{Mitchell:2002wu}, for $|y| < 0.35$ and 
$\Delta\phi = 2 \times \frac{\pi}{2}$, using 2006 data.
Collisions were identified by triggering on a valid 
collision vertex as measured by the beam-beam counters 
(BBC)~\cite{Allen:2003zt}.

Muon track candidates were detected at forward and backward 
rapidities with respect to the polarized beam using two muon 
spectrometers on the north and south sides of the experiment.  
Each spectrometer consists of 3 cathode strip tracking chambers 
in a magnetic field (MuTr) and 5 layers of Iarocci tube planes 
interleaved with thick steel absorbers (MuID).

Muon candidates were selected using the BBC trigger in 
coincidence with combinations of hits in the MuID called 
symsets.  Each symset is defined by projecting from the center 
of the interaction region through a tube in the first layer of 
the MuID to a window of $\pm$3 additional tubes in each 
subsequent layer.  Muon candidates are characterized by the 
depth of their penetration through the MuID.  A deep track 
requires at least one hit of the symset to occur in either the 
first or second layer and one hit in either the fourth or fifth 
layer with at least three layers containing hits, while a 
shallow track requires only one hit in either the first or 
second layer with at least two hits in the first three layers.  
In order for a muon to reach the MuID it must have a minimum 
total momentum of $\sim$1.9~GeV/$c$, and $\sim$2.4~GeV/$c$ is 
needed to reach the last layer.

The dimuon trigger for the 2006 data required at least one muon 
candidate satisfying the deep trigger requirement and another 
satisfying at least the shallow trigger, while the trigger for 
the 2008 data required two deep muon candidates.  After run 
selection, the muon spectrometer on the north~(south) side of 
PHENIX sampled an integrated luminosity of 
1.75~(1.63)~pb$^{-1}$ in 2006 and 4.33~(4.30)~pb$^{-1}$ in 
2008.

For candidate muon pairs, the collision vertex was required to 
be within 35~cm of the center of the interaction region along 
the beam direction, and each track was required to have 
longitudinal momentum $1.4 < p_z $(GeV/$c$) $< 20$.  The 
distance between the track projections from the MuID and the 
MuTr to the first MuID layer was required to be less than 
25~(30)~cm for the detector on the north~(south) side, and the 
angular difference was required to be less than 10~degrees.  A 
fit to the common vertex of the two tracks was performed and 
required to have a $\chi^2<20$ for 4 degrees of freedom.

For an electron candidate at midrapidity, a coincidence was 
required between the BBC trigger and a trigger designed to 
select electrons (ERT), which required a nominal minimum energy 
of 0.4~GeV in a $2 \times 2$-tower region ($\Delta \eta \times 
\Delta \phi = 0.02 \times 0.02$) of the electromagnetic 
calorimeter (EMCal) and corresponding activity in the 
ring-imaging \v{C}erenkov detector (RICH).  After run selection 
the central arm spectrometers sampled an integrated luminosity 
of 1.36~pb$^{-1}$ in 2006.

The momentum of electron candidate tracks was reconstructed 
using the drift chamber (DCH), and their energy was measured 
with the EMCal.  A minimum momentum cut of 0.5~GeV/$c$ was 
applied, and the ratio of the measured energy to momentum 
($E/p$) was required to be within four standard deviations 
($\sigma$) of 1, with the resolution in $E/p$ varying with $p$ 
between 10\% and 15\%.  Position matching between the track in 
the DCH and energy cluster in the EMCal was required to be 
$<4\sigma$ in both the beam direction and azimuth, and only 
collisions occurring within 30~cm of the center of the 
interaction region along the beam direction were considered, 
matching the central arm acceptance.

\subsection{Analysis Method for $J/\psi \rightarrow \mu^{+} 
\mu^{-}$\label{subsec:Method-muon}}

\begin{figure}[t]
\includegraphics[width=0.90\linewidth]{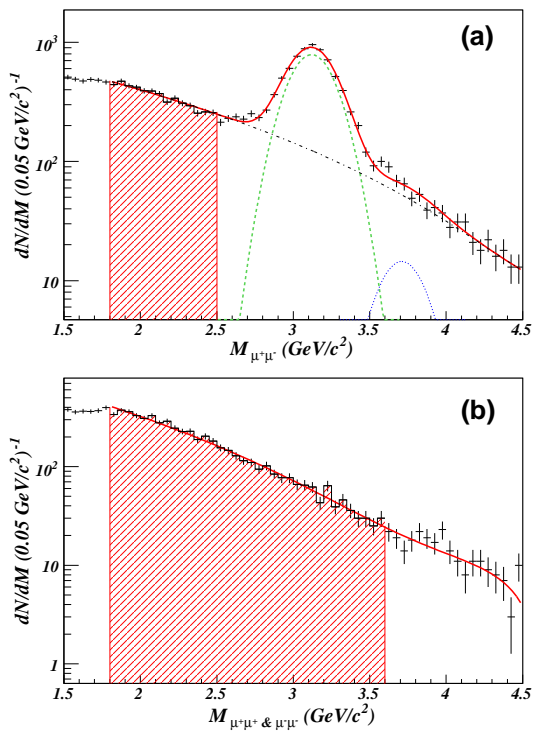}
\caption{(color online).  Invariant mass spectra for (a) 
oppositely-charged muon candidate pairs and (b) charged 
candidate pairs with the same sign for the muon spectrometer on 
the north side of PHENIX from the 2006 data set.  The solid 
line is the sum of the $J/\psi$ (dashed) and 
$\psi^\prime$ (dotted) Gaussians, along with a 
third-order polynomial (dotted-dashed) background.}
\label{fig:mass}
\end{figure}

Transverse single-spin asymmetries for the $J/\psi \rightarrow 
\mu^{+} \mu^{-}$ decay channel were determined by subtracting a 
background asymmetry from the inclusive signal as
\begin{equation}
\label{eq:AN-Phy}
A_N^{J/\psi}=\frac{A_N^{Incl}-r\cdot A_N^{BG}}{1-r},
\end{equation}
where the $A_N$ values on the right-hand-side were calculated 
using Eq.~(\ref{eq:AN-modified}).  The asymmetry $A_N^{Incl}$ 
is for oppositely-charged muon pairs in the invariant mass 
range $\pm 2\sigma$ around the $J/\psi$ mass (where $\sigma$ is 
the mass resolution of the detector), and $A_N^{BG}$ for the 
analysis of the 2006 data set is the asymmetry for 
oppositely-charged muon pairs in the invariant mass range $1.8 
< m$~(GeV/$c^{2}$) $< 2.5$ along with charged pairs of the same 
sign in invariant mass range $1.8 < m$ (GeV/$c^{2}$) $< 3.6$ 
(shaded areas in Fig.~\ref{fig:mass}(a) and (b) respectively).  
For the analysis of the 2008 data set the lower limit of the 
mass range was 2.0~GeV/$c^{2}$.  As a cross check, $A_N^{BG}$ 
was calculated separately for the oppositely-charged muon pairs 
and charged pairs of the same sign, and the results were 
consistent.  Table ~\ref{tab:Bg-asymmetries} gives the 
measured values of $A_N^{BG}$.  The $J/\psi$ mass resolution 
was measured to be 0.143~$\pm$~0.003 
(0.154~$\pm$~0.004)~GeV/$c^2$ in the muon spectrometer on the 
north~(south) side for the 2006 data set and consistent with 
this for the 2008 data set.

\begin{table}[t] \caption{\label{tab:Bg-asymmetries} 
Background asymmetries as a function of $p_{T}$ for PHENIX muon 
spectrometers.  The uncertainties given are statistical.} 
\begin{ruledtabular} \begin{tabular}{cccc} 
$p_T$ (GeV/$c$)& $<x_{F}>$ & data set & $A_N^{BG}$ \\ \hline
0--6           &   -0.081     &     2006    & -0.003$\pm$0.028  \\
               &   -0.082     &     2008    & -0.072$\pm$0.034  \\
               &    0.084     &     2006    & -0.008$\pm$0.028  \\
               &    0.086     &     2008    & -0.003$\pm$0.035  \\ \hline
0--1.4         &   -0.081     &     2006    & -0.021$\pm$0.034  \\
               &   -0.081     &     2008    & -0.089$\pm$0.042  \\
               &    0.085     &     2006    &  0.002$\pm$0.034  \\
               &    0.087     &     2008    & -0.008$\pm$0.043  \\ \hline
1.4--6         &   -0.081     &     2006    &  0.001$\pm$0.053  \\
               &   -0.082     &     2008    & -0.041$\pm$0.068  \\
               &    0.084     &     2006    & -0.039$\pm$0.053  \\
               &    0.086     &     2008    &  0.024$\pm$0.066  \\ 
\end{tabular} \end{ruledtabular}
\end{table}

The background fraction $r$ is defined as
\begin{equation}
\label{eq:bg-fraction}
r=\frac{N_{Incl}-N_{J/\psi}}{N_{Incl}},
\end{equation}
where $N_{Incl}$ is the total number of oppositely-charged muon 
pairs in the invariant mass range $\pm 2\sigma$ around the 
$J/\psi$ mass, and $N_{J/\psi}$ is the $J/\psi$ yield in the 
same mass range.  $N_{J/\psi}$ was extracted by fitting the 
invariant mass spectrum of oppositely-charged muon pairs with 
two Gaussians (for the $J/\psi$ and $\psi^{\prime}$ resonances, 
the peak shape of which is dominated by the detector 
resolution) and a third-order polynomial (for the remaining 
pairs), using the expression:
\begin{eqnarray}
&(a_0&+a_1M+a_2M^2+a_3M^3) \nonumber \\
&+&
\frac{N_{J/\psi}}{2\pi\sqrt{\sigma}}e^{-\frac{(M-M_{J/\psi})^2}{2\sigma^2}}
+\frac{N_{\psi^\prime}}{2\pi\sqrt{\sigma^\prime}}e^{-\frac{(M-M_{\psi^\prime})^2}
{2{\sigma^\prime}^2}}.
\label{eq:bg_shape}
\end{eqnarray}

The free parameters in the fit are the four polynomial 
parameters as well as $N_{J/\psi}$, $M_{J/\psi}$, $\sigma$, and 
$N_{\psi^\prime}$.  The mass and width of the $\psi^\prime$ are 
fixed relative to the $J/\psi$ based on simulations.  The 
polynomial is used to fit the background from both physical 
sources (i.e.  Drell-Yan, open heavy flavor) and uncorrelated 
track combinations.  Uncorrelated track combinations comprise 
more than 50\% of the oppositely-charged muon pairs under the 
$J/\psi$ mass peak.  An example mass spectrum and fit is shown 
in Fig.~\ref{fig:mass}(a).  The total $J/\psi$ yield in the 
dimuon decay channel was 6403~$\pm$~126 for the 2006 and 
15380~$\pm$~150 for the 2008 data set.  
Table~\ref{tab:Bg-fraction} gives background fractions in 
different $p_T$ ranges determined from the fit.  The asymmetries 
were measured store-by-store, using Eq.~(\ref{eq:AN-modified}), 
with the final results obtained by averaging over all stores.

A geometric scale factor from 2006 data of $f=1.57\pm0.04$ was 
determined from $J/\psi$ azimuthal distributions in data and 
was found to be independent of $p_{T}$ within statistical 
uncertainties.  In the 2008 data, the polarization of the 
clockwise circulating beam was found to be rotated by 
0.25~$\pm$~0.033 rad away from vertical (see 
Appendix~\ref{sec:local_pol}), meaning that $f$ was different 
depending on which beam was considered polarized in the 
analysis.  The geometric scale factors from that analysis were 
$f=1.64\pm0.01$ for the clockwise circulating beam and 
$f=1.56\pm0.01$ for the counter-clockwise circulating beam

\begin{table}[t]
\centering \caption{\label{tab:Bg-fraction}
Total background fractions as a function of $p_{T}$ for muon 
spectrometers on the north and south sides of PHENIX.  
Backgrounds were higher in the 2006 data set because the less 
restrictive trigger requirement allowed more random track 
combinations.}
\begin{ruledtabular} \begin{tabular}{ccccc}
$p_T$ (GeV/$c$)   &    data set   & \ \ detector \ \  & background fraction (\%)\\ \hline
0--6              &     2006     &       South       & 21.7$\pm$0.6    \\
                  &     2006     &       North       & 19.1$\pm$0.4    \\
                  &     2008     &       South       & 16.4$\pm$0.2    \\
                  &     2008     &       North       & 14.2$\pm$0.2    \\ \hline
0--1.4            &     2006     &       South       & 23.2$\pm$0.7    \\
                  &     2006     &       North       & 22.0$\pm$0.7    \\
                  &     2008     &       South       & 16.1$\pm$0.3    \\
                  &     2008     &       North       & 15.5$\pm$0.3    \\ \hline
1.4--6            &     2006     &       South       & 20.1$\pm$0.8    \\
                  &     2006     &       North       & 14.1$\pm$0.5    \\
                  &     2008     &       South       & 15.6$\pm$0.4    \\
                  &     2008     &       North       & 10.5$\pm$0.2    \\
\end{tabular} \end{ruledtabular}
\end{table}

Several systematic checks were performed in the analysis.  
Asymmetries were determined using random spin orientations to 
cross check their consistency with zero.  The means were found 
consistent with zero and widths consistent with statistical 
uncertainties in $A_N$.  The parity-odd asymmetries along the 
proton spin direction were also measured and found to be 
consistent with zero as expected for the strong interaction.

Following previous PHENIX publications systematic uncertainties 
in this analysis are categorized as Type A, point-to-point 
uncorrelated, Type B, scaling all points in the same direction 
but not by the same factor, and Type C, scaling all points by 
the same factor.  It should be noted that all scale 
uncertainties affect both the central values and the 
statistical uncertainties such that the statistical 
significance of the measurement from zero is preserved.  A 
single Type A systematic uncertainty is included, which is due 
to the fit used to determine the background.  The fit was 
nominally performed using a range of $1.8 < m $ (GeV/$c^{2}$) 
$< 5.0$ for the 2006 and $2.0 < m $ (GeV/$c^{2}$) $<5.0$ for 
the 2008 data.  To estimate the uncertainty due to the choice of 
fit range, the fit was then performed using a range of $1.5 < m 
$ (GeV/$c^{2}$) $< 5.5$, and the difference of the calculated 
$A_N$ with the nominal was taken as a systematic uncertainty.  
Other systematic uncertainties are discussed in 
Section~\ref{sec:Result}.

\subsection{Analysis Method for $J/\psi \rightarrow e^{+} e^{-}$\label{subsec:Method-central}}

For the RHIC luminosities and store lengths in 2006 there were 
too few $J/\psi$s detected in the PHENIX central arms to 
calculate a separate asymmetry for each store.  Instead, 
asymmetries were calculated using Eq.~(\ref{eq:AN-sqrt}) from 
statistics integrated across the four-week running period.  
The integration of statistics over a long time-period requires 
special care to be taken in order to avoid introducing false 
asymmetries.

If we measure an asymmetry in $n$ time periods with, for 
simplicity, equivalent statistics in each measurement $i$, we 
can determine an averaged asymmetry in the left hemisphere 
using Eq.~(\ref{eq:AN-lumi}) as
\begin{equation}
A_{N} = \frac{1}{n} \displaystyle\sum_{i=1}^{n} \frac{f_{i}}{\mathcal{P}_{i}} \frac{N_{i}^{\uparrow} - \mathcal{R}_{i} N_{i}^{\downarrow}}{N_{i}^{\uparrow} + \mathcal{R}_{i} N_{i}^{\downarrow}} = \frac{1}{n} \displaystyle\sum_{i=1}^{n} A_{N,i}.
\end{equation}
Since each measurement is probing the same physical observable we have the same $A_{N,i}$ for all $i$ within statistical uncertainties.  This allows us to calculate the asymmetry as
\begin{equation}
\displaystyle A_{N} = \frac{\displaystyle\sum_{i=1}^{n}f_{i} \left(N_{i}^{\uparrow} - \mathcal{R}_{i} N_{i}^{\downarrow}\right)}{\displaystyle\sum_{i=1}^{n} \mathcal{P}_{i} \left(N_{i}^{\uparrow} + \mathcal{R}_{i} N_{i}^{\downarrow}\right)},
\end{equation}
which corresponds to integrating our $n$ measurement periods into a single measurement.  For a statistically limited measurement we do not calculate $\mathcal{R}_{i}$, $f_{i}$, and $\mathcal{P}_{i}$ for all $i$ but assume that the measurement can be made as
\begin{equation}
\displaystyle A_{N} = \frac{\left<f\right>\displaystyle\sum_{i=1}^{n} \left( N_{i}^{\uparrow} - \left<\mathcal{R}\right> N_{i}^{\downarrow} \right)}{\left<\mathcal{P}\right>\displaystyle\sum_{i=1}^{n}\left( N_{i}^{\uparrow} + \left<\mathcal{R}\right> N_{i}^{\downarrow} \right)},
\end{equation}
where the brackets denote a luminosity-weighted average over 
the course of the measurement.  In order for the same physical 
observable to be calculated by this expression without 
additional systematic uncertainty we must have $\mathcal{R}_{i} 
= \left<\mathcal{R}\right>$, $f_{i}=\left<f\right>$, and 
$\mathcal{P}_{i}=\left<\mathcal{P}\right>$ for all $i$.

The PHENIX central arm acceptance was stable enough over the 
course of the running period so that the assumption is valid 
for $f$, but variations in the polarization and relative 
luminosity must be taken into account.  Variations in the 
polarization were found to be consistent with statistical and 
systematic errors on measurements of the polarization and 
contribute to an uncertainty in the overall scale of $A_N$ 
(scale uncertainties will be discussed in more detail in 
Section~\ref{sec:Result}).  Variations in $\mathcal{R}$ have 
the most significant effect on the asymmetry as they can 
potentially contribute to false asymmetries.

In order to stabilize relative luminosities a procedure was 
developed wherein several bunches of colliding protons are 
removed from the analysis so that the relative luminosity 
within each store is brought as close to unity as possible 
given the finite number of bunches.  In principle we could 
choose any constant value for the relative luminosity, but 
unity is chosen for convenience.

First, all bunches with luminosities greater than 2 standard 
deviations away from the mean bunch luminosity of a store are 
removed.  A bunch is then chosen at random, and if removing 
this bunch from the analysis brings the relative luminosity 
closer to unity it is removed.  Otherwise it is kept, and 
another bunch is chosen at random.  The process continues until 
the relative luminosity is within 1$\%$ of unity or as close to 
unity as possible given the finite number of bunches.  The 
corrected relative luminosities were distributed with an RMS of 
approximately 1.5$\%$ away from unity, and the entire procedure 
removed approximately 5$\%$ of the provided luminosity from the 
data sample used in this analysis.

Since removed bunches are chosen at random, the result of the 
analysis is not unique, and a systematic uncertainty is 
introduced.  In order to determine this systematic uncertainty, 
the analysis was run 5000 times, and the resulting asymmetries 
and statistical uncertainties were histogrammed.  The mean 
values of these two histograms were then taken as the central 
value and statistical uncertainty, and the RMS of the 
histogrammed asymmetry was taken as the dominant systematic 
uncertainty in the analysis.

It is not generally true that detector efficiencies are 
independent of spin orientation in the PHENIX central arms when 
triggering with the ERT, as even and odd numbered crossings are 
triggered by separate circuits with slightly different gains.  
While the results for even and odd crossings are consistent 
within statistical uncertainties and no systematic trend is 
observed, to eliminate any possible effects from the different 
trigger circuits we measure asymmetries separately for even and 
odd crossings and combine the resulting asymmetries.

\begin{figure}[th]
\includegraphics[width=1.0\linewidth]{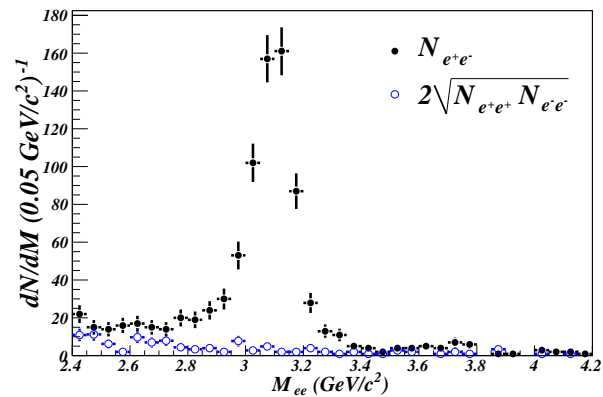}
\caption{(color online).  Invariant mass spectra for 
oppositely-charged electron candidate pairs and uncorrelated 
track pairs defined as $2\sqrt{N_{e^{+}e^{+}}N_{e^{-}e^{-}}}$ 
in the central spectrometer.}
\label{fig:mass_central}
\end{figure}

The PHENIX central arms detected 539~$\pm$~25 $J/\psi$ mesons 
in an invariant mass range $2.7 < m $ (GeV/$c^{2}$) $< 3.4$.  
The resolution of the $J/\psi$ mass peak was found to be 
0.061~$\pm$~0.002~GeV/$c^{2}$.  Charged lepton pairs with the 
same sign are due to uncorrelated tracks and are subtracted 
from the oppositely-charged pairs using the number of 
like-signed pairs as $2\sqrt{N_{e^{+}e^{+}}N_{e^{-}e^{-}}}$.  
Both mass spectra can be found in Fig.~\ref{fig:mass_central}.  
After subtracting the uncorrelated tracks from the mass 
spectra, there is a remaining continuum background from open 
heavy flavor and Drell-Yan production as well as a small number 
of lepton pairs from $\psi^{\prime}$ decays reconstructed to 
low mass.  The contribution from conversion electrons is 
negligible.  The background fraction from continuum pairs as 
defined in Eq.~(\ref{eq:bg-fraction}) can be found in 
Table~\ref{tab:Bg-fraction-central}.  There were not enough 
statistics in the continuum background to determine $A_N^{BG}$, 
and inclusion of such background was not found to significantly 
affect the signal asymmetry.  An overall dilution of the signal 
is included assuming that $A_N^{BG}=$0.

\begin{table}[t]
\centering \caption{\label{tab:Bg-fraction-central}  Fraction of measured electron pairs coming from the continuum background in the central spectrometer.}
\begin{ruledtabular} \begin{tabular}{cccc}
& $p_T$ (GeV/$c$) &  $r$ (\%)    & \\ \hline
& 0--6            & 6.6$\pm$0.4  &  \\ 
& 0--1.4          & 5.6$\pm$0.5  &  \\ 
& 1.4--6          & 7.8$\pm$0.7  &  \\ 
\end{tabular} \end{ruledtabular}
\end{table}

\begin{table}[h]
\centering \caption{\label{tab:central-fs} 
Geometric scale factors determined from simulation for $J/\psi 
\rightarrow e^{+}e^{-}$.  }
\begin{ruledtabular} \begin{tabular}{cccc}
& $p_{T}$ (GeV/$c$) & $f^{\prime}$    & \\ \hline
& 0--6              & 1.62$\pm$0.01   & \\ 
& 0--1.4            & 1.61$\pm$0.01   & \\
& 1.4--6            & 1.70$\pm$0.02   & \\ 
\end{tabular} \end{ruledtabular}
\end{table}

There were not enough $J/\psi$s detected in order to determine 
geometric scale factors for the asymmetries at midrapidity from 
data.  These factors were instead determined using a 
GEANT~\cite{GEANT} Monte Carlo simulation of single $J/\psi$ 
decays with a full geometric description of the central arm 
detectors including all known inefficiencies.  Geometric scale 
factors used in the analysis are given in 
Table~\ref{tab:central-fs} and mean polarizations in 
Section~\ref{subsec:SSA}.

A number of systematic checks were performed to ensure the 
validity of the measurement.  The number of $J/\psi$s produced 
divided by the provided beam luminosity was calculated and 
found to be constant across stores.  Asymmetries were 
determined with randomized spin directions as described in 
Section~\ref{subsec:Method-muon} and found to be distributed in 
normal distributions which had means consistent with zero and 
widths consistent with the statistical uncertainties.  
Parity-odd asymmetries along the proton spin direction were 
measured and found to be consistent with zero, as expected.

\section{Results and summary\label{sec:Result}}

\begin{figure}[htbp]
\includegraphics[width=1.0\linewidth]{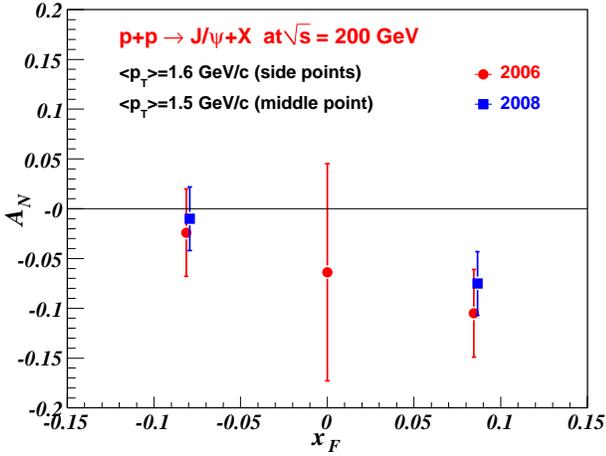}
\caption{(color online) Transverse single-spin asymmetry in 
$J/\psi$ production as a function of $x_F$ for 2006 and 2008 
data sets.  The error bars shown are statistical and Type A 
systematic uncertainties added in quadrature.  Type B 
systematic uncertainties are not included but are 0.003 or less 
in absolute magnitude and can be found in Table~\ref{tab:AN}.  
An additional uncertainty in the scale of the ordinate due to 
correlated polarization uncertainties of 3.4\% (3.0\%) for the 
2006 (2008) data set is not shown.  See text for details.} 
\label{fig:AN-xF-run68}
\end{figure}

\begin{figure}[htbp]
\includegraphics[width=1.0\linewidth]{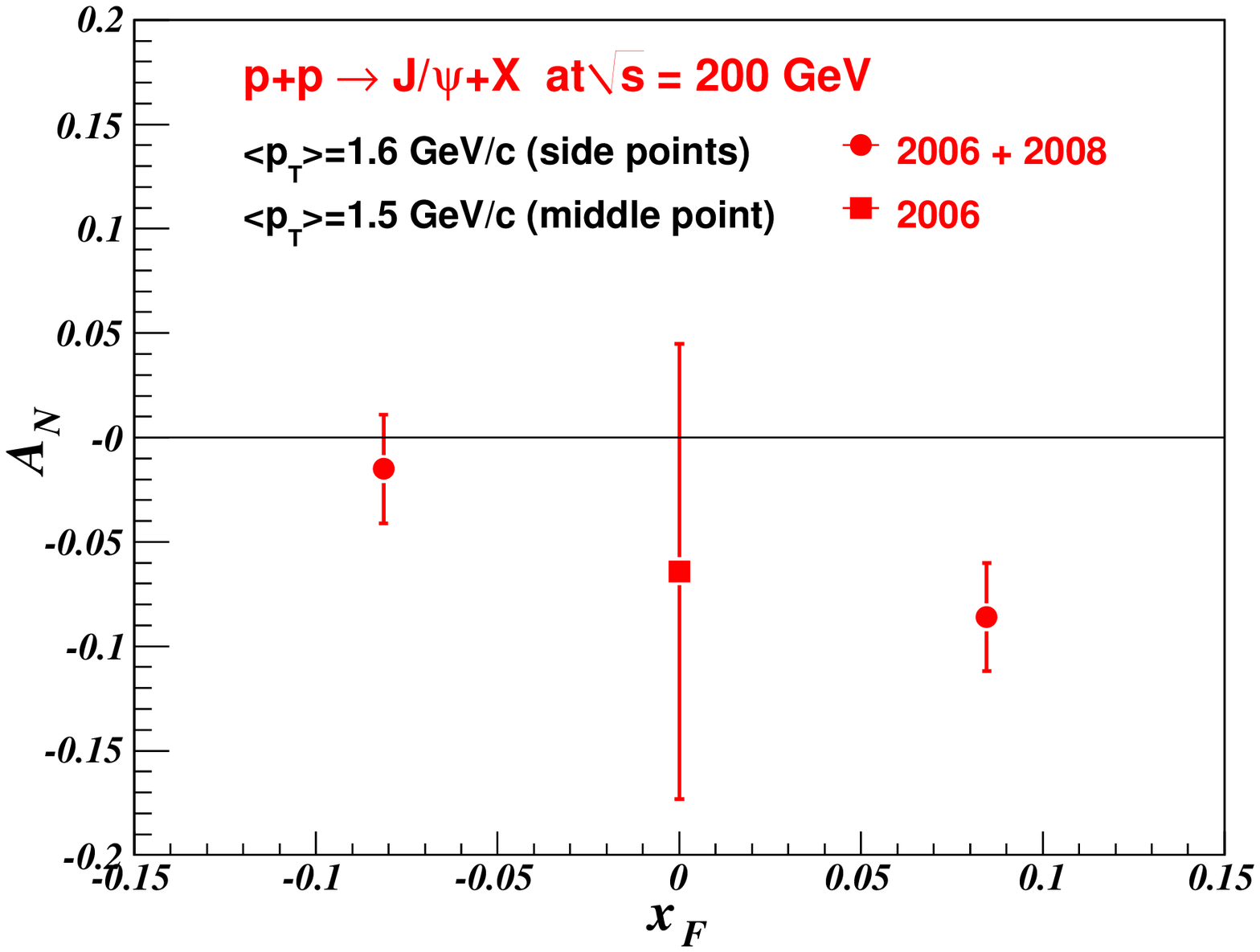}
\caption{(color online) Transverse single-spin asymmetry in 
$J/\psi$ production as a function of $x_F$ for combined 2006 
and 2008 data sets.  The error bars shown are statistical and 
Type A systematic uncertainties added in quadrature.  Type B 
systematic uncertainties are not included but are 0.002 or less 
in absolute magnitude and can be found in Table~\ref{tab:AN}.  
An additional uncertainty in the scale of the ordinate due to 
correlated polarization uncertainties of 2.4\% (3.4\%) for the 
points with $|x_F|>0$ ($x_F=0$) is not shown.  See text for 
details.} \label{fig:AN-xF}
\end{figure}

\begin{figure}[th]
\includegraphics[width=1.0\linewidth]{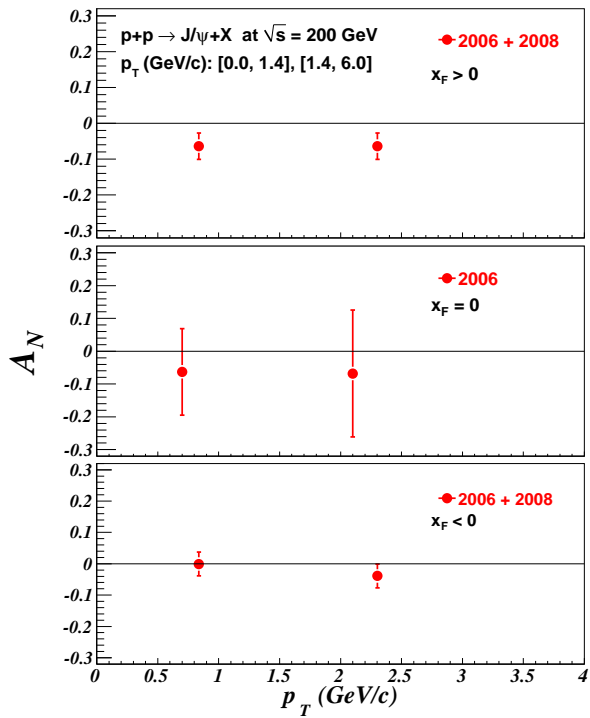}
\caption{(color online) Transverse single-spin asymmetry of 
$J/\psi$ mesons plotted against $J/\psi$ transverse momentum.  
See Table~\ref{tab:AN} for mean $x_F$ values for each point.  
The error bars shown are statistical and Type A systematic 
uncertainties added in quadrature.  Type B systematic 
uncertainties are not included but are 0.002 or less in 
absolute magnitude and can be found in Table~\ref{tab:AN}.  An 
additional uncertainty in the scale of the ordinate due to 
correlated polarization uncertainties of 2.4\% (3.4\%) for the 
points with $|x_F|>0$ ($x_F=0$) is not shown.  See text for 
details.}
\label{fig:AN-pT}
\end{figure}

Figures~\ref{fig:AN-xF-run68} and~\ref{fig:AN-xF} present the 
measured transverse single-spin asymmetry in $J/\psi$ 
production versus $x_F$, and Fig.~\ref{fig:AN-pT} shows the 
measured transverse SSA at different rapidities as a function 
of $p_T$.  The results are tabulated in Table~\ref{tab:AN}.  
The Type A systematic uncertainties, which are point-to-point 
uncorrelated, are due to the procedure used to stabilize the 
relative luminosity in the midrapidity analysis, as described 
in Section~\ref{subsec:Method-central}, and the fit used to 
determine the background in the analysis at forward and 
backward rapidities, as described in 
Section~\ref{subsec:Method-muon}.  As discussed in 
Section~\ref{subsec:SSA}, uncertainties in the geometric scale 
factors and polarizations lead to a fractional scale 
uncertainty on $A_N$.  The uncertainties shown in the last two 
columns of Table~\ref{tab:AN} are Type B systematic 
uncertainties, scaling all $x_F$ or $p_T$ points in the same 
data set in the same direction but not necessarily by the same 
factor.  There are additional Type C scale uncertainties, which 
scale all points in exactly the same way, due to the fully 
correlated polarization uncertainties in each data set of 3.4\% 
for 2006, 3.0\% for 2008, and 2.4\% for the combined 2006 and 
2008 data sets.  It should be noted that all scale uncertainties 
affect both the central values and the statistical 
uncertainties such that the statistical significance of the 
measurement from zero is preserved.  While in principle the 
polarization uncertainties do not affect $A_N$ symmetrically 
due to the fact that $A_N \propto \frac{1}{P}$, the difference 
in the value of the uncertainties scaling to larger and smaller 
magnitudes of $A_N$ is less than the precision shown.

As the functional form of the asymmetry in $x_F$ and $p_T$ is 
completely unknown, no correction has been made for potential 
smearing effects.  A simulation study was performed assuming a 
linear dependence of $A_N$ on $x_F$, and it was found that 
smearing effects were less than 10\% of the value of the input 
asymmetry.

\begin{table*}[th]
\caption{\label{tab:AN}
$A_N$ vs.  $p_T$ in forward, backward and midrapidity.  
Systematic uncertainties in the last two columns are due to the 
geometric scale factor and the polarization, respectively.  
There are additional Type C uncertainties due to the 
polarization of 3.4\%, 3.0\%, and 2.4\% for the 2006, 2008, and 
combined 2006 and 2008 results.  See text for details.}
\begin{ruledtabular} \begin{tabular}{cccccccc}
 $p_T$      & Data Sample& $<x_F>$ & $A_N$& $\delta A_N$ & $\delta A_N$ & $\delta A_N^f$ (\%) & $\delta A_N^P$ (\%) \\
(GeV/$c$)     &            &          &        & (stat)  & (Type A syst)& (Type B syst)      &  (Type B syst.)     \\ \hline
           & 2006        & -0.081   & -0.024  & 0.044   & 0.003         & 0.6                 & 2.3 \\
           & 2008        & -0.082   & -0.010  & 0.032   & 0.004         & 0.4                 & 3.4 \\
           & 2006 + 2008 & -0.081   & -0.015  & 0.026   & 0.002         & 0.4                 & 2.8 \\
           \cline{2-8}

0--6       & 2006        &  0.000   & -0.064  & 0.106   & 0.026         & 0.6                 & 2.3 \\
\cline{2-8}

           & 2006        &  0.084   & -0.105  & 0.044   & 0.005         & 0.6                 & 2.3 \\
           & 2008        &  0.086   & -0.075  & 0.032   & 0.003         & 0.4                 & 3.3 \\
           & 2006 + 2008 &  0.085   & -0.086  & 0.026   & 0.003         & 0.4                 & 2.7 \\
           \hline

            & 2006        & -0.081   &  0.050  & 0.067  & 0.007         & 0.6                 & 2.3 \\
            & 2008        & -0.081   & -0.025  & 0.046  & 0.008         & 0.4                 & 3.4 \\
            & 2006 + 2008 & -0.081   & -0.001  & 0.038  & 0.005         & 0.4                 & 2.8 \\
            \cline{2-8}

0--1.4      & 2006        & 0.000    & -0.063  & 0.128  & 0.031         & 0.6                 & 2.3\\
\cline{2-8}

            & 2006        & 0.085   & -0.065  & 0.066   & 0.005         & 0.6                 & 2.3 \\
            & 2008        & 0.087   & -0.064  & 0.045   & 0.003         & 0.4                 & 3.4 \\
            & 2006 + 2008 & 0.086   & -0.064  & 0.037   & 0.003         & 0.4                 & 2.7 \\
            \hline

            & 2006        & -0.081   & -0.073  & 0.065 & 0.002          & 0.6                 & 2.3 \\
            & 2008        & -0.082   & -0.023  & 0.046 & 0.010          & 0.4                 & 3.5 \\
            & 2006 + 2008 & -0.082   & -0.039  & 0.038 & 0.002          & 0.4                 & 2.8 \\
            \cline{2-8}

1.4--6      & 2006        & 0.000   & -0.068  & 0.188  & 0.045          & 1.2                 & 2.3 \\
\cline{2-8}

            & 2006        & 0.084   & -0.046  & 0.064  & 0.005          & 0.6                 & 2.3 \\
            & 2008        & 0.086   & -0.073  & 0.046  & 0.007          & 0.4                 & 3.3 \\
            & 2006 + 2008 & 0.085   & -0.064  & 0.037  & 0.004          & 0.4                 & 2.7 \\
\end{tabular} \end{ruledtabular}
\end{table*}

The measured asymmetry at forward $x_F$ is negative, -$0.086 \pm 
0.026 \pm 0.003$, with a statistical significance from zero of 
$3.3\sigma$, suggesting a nonzero trigluon correlation function in 
transversely polarized protons and, if well defined as a universal 
function in the reaction $p+p \rightarrow J/\psi + X$, a nonzero 
gluon Sivers function.  A nonzero transverse SSA in $J/\psi$ 
production in $p+p$ generated by gluon dynamics may seem surprising 
given the SSAs consistent with zero in midrapidity neutral pion 
production at PHENIX~\cite{Adler:2005in} and semi-inclusive charged 
hadron production at COMPASS~\cite{Alexakhin:2005iw}.  However, the 
details of color interactions have been shown to play a major role 
in SSAs~\cite{Rogers:2010dm}, so further theoretical development 
will be necessary before we fully understand the relationships among 
these measured asymmetries.  As discussed in 
Ref.~\cite{Yuan:2008vn}, a nonzero transverse SSA in $J/\psi$ 
production in polarized $p+p$ collisions generated by a 
gluon Sivers TMD would be evidence against large contributions from 
color-octet diagrams for $J/\psi$ production.  If a gluon Sivers TMD 
is in fact well defined and nonzero, a new experimental avenue has 
been opened up to probe the $J/\psi$ production mechanism, a 
long-standing question in QCD.  As discussed in 
Section~\ref{sec:Introduction}, there is a relationship between the 
SSA and the production mechanism of the $J/\psi$ in the collinear, 
higher-twist approach, but it is not as simple as in the TMD 
approach.

Future $p+p$ data from RHIC are expected to improve the precision of 
the current measurement, and a similar measurement for $J/\psi$ 
production in SIDIS with a transversely polarized target could shed 
further light on the production mechanism, as \cite{Yuan:2008vn} 
predicts a vanishing asymmetry for the color-singlet model in SIDIS 
but nonzero asymmetry for the color-octet model.  While no rigorous 
quantitative calculations are presently available for either 
collision system, we anticipate that future theoretical calculations 
will provide more detailed guidance on the implications of the 
present results.


\section*{ACKNOWLEDGMENTS}


We thank the staff of the Collider-Accelerator and Physics
Departments at Brookhaven National Laboratory and the staff of
the other PHENIX participating institutions for their vital
contributions.  
We also thank Feng Yuan, Jianwei Qiu, and Zhongbo
Kang for helpful discussions.
We acknowledge support from the Office of Nuclear Physics in the
Office of Science of the Department of Energy, 
the National Science Foundation,  
a sponsored research grant from Renaissance Technologies LLC, 
Abilene Christian University Research Council, 
Research Foundation of SUNY, 
and Dean of the College of Arts and Sciences, Vanderbilt University 
(USA),
Ministry of Education, Culture, Sports, Science, and Technology
and the Japan Society for the Promotion of Science (Japan),
Conselho Nacional de Desenvolvimento Cient\'{\i}fico e
Tecnol{\'o}gico and Funda\c c{\~a}o de Amparo {\`a} Pesquisa do
Estado de S{\~a}o Paulo (Brazil),
Natural Science Foundation of China (People's Republic of China),
Ministry of Education, Youth and Sports (Czech Republic),
Centre National de la Recherche Scientifique, Commissariat
{\`a} l'{\'E}nergie Atomique, and Institut National de Physique
Nucl{\'e}aire et de Physique des Particules (France),
Ministry of Industry, Science and Tekhnologies,
Bundesministerium f\"ur Bildung und Forschung, Deutscher
Akademischer Austausch Dienst, and Alexander von Humboldt Stiftung (Germany),
Hungarian National Science Fund, OTKA (Hungary), 
Department of Atomic Energy (India), 
Israel Science Foundation (Israel), 
National Research Foundation and WCU program of the 
Ministry Education Science and Technology (Korea),
Ministry of Education and Science, Russia Academy of Sciences,
Federal Agency of Atomic Energy (Russia),
VR and the Wallenberg Foundation (Sweden), 
the U.S.  Civilian Research and Development Foundation for the
Independent States of the Former Soviet Union, 
the US-Hungarian Fulbright Foundation for Educational Exchange,
and the US-Israel Binational Science Foundation.

\appendix*

\section{Local Polarimetry} \label{sec:local_pol}

The transverse component of the proton polarization is
monitored at PHENIX by a measurement of the SSA of forward
neutron production with a Zero-Degree Calorimeter (ZDC) and
Shower Maximum Detector (SMD), collectively referred to as
the local polarimeter.

The ZDC consists of three modules of hadronic calorimeter, 1.7 interaction length each, located approximately 18~m from the center of the interaction region.  The SMD consists of two layers of plastic scintillator arrays located between the first and second modules of the ZDC, and it provides horizontal and vertical position
information of the detected neutrons.  There is an additional plastic
scintillator in front of the ZDC used to identify and veto charged particles.

Neutron candidates were selected by triggering on a coincidence between the
BBC trigger and a signal in either the north or south ZDC.  Due to
 bandwidth restrictions, the trigger was prescaled by a factor typically around 100.
The analysis of the 2006 data used $8.7~\times~10^7$ events from radially polarized 
collisions, and the analysis of the 2008 data used $7.3~\times~10^7$ events from 
vertically polarized collisions.

Neutrons were selected offline by requiring an energy deposit in the ZDC between 20 and 120~GeV along with no hit in the veto scintillator (less than 1 minimum-ionizing particle).  Hits were required in both the horizontal and vertical planes of the SMD, and neutrons were required to be between 0.3~mrad (0.5~cm at the SMD) and 1.4~mrad (2.5~cm at the SMD) from the beam.

The asymmetry is calculated using fits to the azimuthal distribution of the neutrons:
\begin{eqnarray*}
& A(\phi) = A_N \cos(\phi - \phi_0) & {\rm (2006, radially~polarized)}  \\
& A(\phi) = A_N \sin(\phi - \phi_0) & {\rm (2008, vertically~polarized)}
\end{eqnarray*}
where $\phi$ is the azimuthal angle relative to vertical, and 
$\phi_0$ is the deviation of the polarization from the nominal 
direction.

The position resolution of the local polarimeter is approximately 
1~cm.  To determine a systematic uncertainty on the polarization 
direction, fits are performed for $\pm$1~cm around the center of the 
azimuthal neutron distribution (defined as the position where $A_N$ 
is maximal).  The deviation of the polarization from nominal, in 
radians, was found to be
\begin{eqnarray*}
& 0.064 \pm 0.040^{\rm stat} \pm 0.086^{\rm syst} & {\rm (clockwise)} \\
& 0.109 \pm 0.038^{\rm stat} \pm 0.036^{\rm syst} & {\rm (counterclockwise)}
\end{eqnarray*}
for the 2006 data and
\begin{eqnarray*}
& 0.263 \pm 0.030^{\rm stat} \pm 0.090^{\rm syst} & {\rm (clockwise)} \\
& 0.019 \pm 0.048^{\rm stat} \pm 0.103^{\rm syst} & {\rm (couterclockwise)}
\end{eqnarray*}
for the 2008 data.  The polarization directions of the 
counterclockwise-going beam in 2006 and clockwise-going beam in 2008 
were considered to be significant and used in the calculation of 
$A_N$ for $J/\psi$.


\end{document}